\documentclass[twocolumn,showpacs,preprintnumbers,amsmath,amssymb]{revtex4}

\usepackage[T1]{fontenc}
\usepackage{graphicx}
\usepackage{dcolumn}
\usepackage{bm}

\usepackage[ansinew]{inputenc} 
\usepackage{subfigure} 
\usepackage{lscape} 
\usepackage{amstext} 

\usepackage{color,ifpdf,graphicx}
\ifpdf 
  \DeclareGraphicsExtensions{.pdf,.png,.mps}
\else 
  \DeclareGraphicsExtensions{.eps,.bmp}
  \usepackage{pstricks}
\fi
\usepackage{epic,bez123,wrapfig}

\makeindex

\begin{document}

\title{$\Lambda$ hyperonic effect on the magnesium dripline}

\author{Torsten Schürhoff}
\affiliation{Frankfurt Institute for Advanced Studies(FIAS),
Institute for Theoretical Physics(ITP), Johann Wolfgang Goethe
University, Frankfurt am Main, Germany}

\author{Stefan Schramm}
\affiliation{Frankfurt Institute for Advanced Studies(FIAS),
Institute for Theoretical Physics(ITP), Center for Scientific
Computing(CSC), Johann Wolfgang Goethe University, Frankfurt am
Main, Germany}

\author{Chhanda Samanta}
\affiliation{Department of Physics and Astronomy, Virginia
Military Institute, Lexington, VA 24450, USA}

\date{\today}

\begin{abstract}

Neutron dripline calculations for both magnesium nuclei and
magnesium + $\Lambda$ hypernuclei have been carried out in a
microscopic framework using a chiral effective model. The
results are compared with two other relativistic mean field models, SPL-40 and NL3. All
three models describe the $\Lambda$ separation energy of known
hypernuclei adequately. The extrapolation to the driplines for
moderately heavy hypernuclei are found to be strongly model-dependent.
\end{abstract}

\maketitle

\section{Introduction}

Hypernuclei are of particular interest in nuclear physics and
nuclear astrophysics. Hyperons, consisting of up, down and strange
quarks, have a very short life time on the weak-interacton timescale ($\sim 10^{-10}s$) and can
form a bound system with nucleons. Thus they can supply the most
direct information on the interaction of hyperons in nuclear
matter, which is also essential for the study and possible masses of
neutron stars.

The hyperon is not Pauli-blocked by the nucleons and thus can occupy the lowest energy level
while providing extra binding.  This can make an unbound normal
nucleus bound (viz., the nucleus $^{9}Li+n$ is unbound, but
$^{9}Li+\Lambda$ is bound), and may even lead to the production of
exotic nuclei beyond the normal drip lines. Hence hypernuclei are
an experimental testing ground to study how a system of up- and
down-quarks changes towards the full flavor SU(3), when strange
quarks are added implicitly through hyperons.

The binding energy of various hypernuclei has already been measured. 
To increase this number, experimental
searches for neutron rich hypernuclei have been carried out at
JPARC, Japan \cite{Sugimura2014}, \cite{Sugimura2013}, and
JLab, USA \cite{Gogami2013}, \cite{Nakamura2011}.
  Also, there are proposed experiments like HypHI
\cite{HypHI}, \cite{HypH1Results} at GSI, Darmstadt, Germany
and NICA~ \cite{Botvina} in Dubna, Russia. A review of some of
the future research activities can be found in \cite{NextDecade}. These  proposed experiments will probe
higher-mass hypernuclei and investigate neutron-rich hyper nuclei towards the hypernuclear neutron
dripline, similar to investigations into the nuclear dripline.

In this context we present a study of the position of the drip
line of magnesium isotopes and consider the impact of an
additional $\Lambda$ hyperon on neutron-rich isotopes. The
calculations are based on a new parameterization of a
well-established SU(3) chiral effective approach
 for studying hadronic matter and nuclei.
The results are compared to two different relativistic
Walecka-type models, SPL-40 \cite{PL40Original},\cite{PL40} and NL3 \cite{NL3}.

The outline of the paper is as following: in section 2
the chiral model used for our calculation is described. In section 3 we show
the results for the dripline calculations. In section 4
the results are discussed and we give a conclusion.

\section{Description of the model}

The Frankfurt chiral model uses an effective SU(3) sigma-omega
model approach to describe nuclear physics. A first version of the
model was developed in \cite{Papazoglou 98a} \cite{Papazoglou
Paper 98} and has been stepwise expanded to describe cross-checks
with heavy-ion experiments \cite{Zschiesche2001}, the inclusion of higher
baryon resonances \cite{Baryon Resonances Zschiesche} and
neutron star physics \cite{VeronicaHybrid}.

In a paper preceding this work \cite{SWSDeformedNuclei}, nuclear
binding energies were fitted. While this approach gives good
results for nuclear binding energies, it also produces a maximum
neutron star mass of just 1.64 $M_{\odot}$. More recent neutron star
mass measurements by Demorest et al. \cite{Demorest} and
Antoniadis et al. \cite{2Solar2013} point to a maximum mass of
neutron stars of at least 1.97 $\pm$ 0.04 $M_{\odot}$ or 2.01
$\pm$ 0.04 $M_{\odot}$.

In order to keep the model up-to-date, certain key parameters have
been reexamined. Among them are the nucleon-omega meson coupling
$(g_{N\omega})$, the nucleon-rho meson coupling $g_{N\rho}$, the
vector-meson self-interaction $g_4^4$ and several other model parameters. 

The degrees
of freedom included in the model are the baryon octet (n,p, and hyperons $\Lambda,
\Sigma^{+,0,-}, \Xi^{0,-}$) and the leptons(e, $\mu$).
The model uses an effective approach where the mesons
mediate the interactions between the baryons. The relevant mesonic degrees of freedom in the mean-field
calculation preformed in these studies are the
vector-isoscalar $\omega$ and $\phi$, the vector-isovector $\rho$
and the scalar-isoscalar $\sigma$ and $\zeta$ (strange
quark-antiquark state).

The Lagrangian density contains the following terms

\begin{eqnarray}
&L = L_{Kin}+L_{Int}+L_{Self}+L_{SB},&
\end{eqnarray}

\noindent where $L_{Kin}$ is the kinetic energy term for the hadrons.
In addition there is an interaction term between the baryons and
the scalar and vector mesons

\begin{eqnarray}
&L_{Int}=-\sum_i
\bar{\psi_i}[\gamma_0(g_{i\omega}\omega+g_{i\phi}\phi+g_{i\rho}\tau_3\rho)+M_i^*]\psi_i\nonumber,&\\&
\end{eqnarray}
with the effective mass $M_i^*$ given by
\begin{eqnarray}
&M_i^* = g_{i\sigma} \sigma + g_{i\zeta} \zeta + \delta m_i  .&
\end{eqnarray}
with a small bare mass term $\delta m_i$. The coupling strengths
of the baryons to the scalar fields are connected via SU(3)
symmetry relations and the different SU(3) invariant coupling
strengths are fitted to reproduce the baryon masses in vacuum (see
ref. \cite{VeronicaHybrid} for the values of the couplings).

\noindent The self-interaction terms for the scalar  mesons read

\begin{eqnarray}\label{ScalarPotentialeq}
\mathcal{L}_{Scalar}&=&\frac{1}{2}k_0\chi^2(\sigma^2+\zeta^2+\delta^2)-k_1(\sigma^2+\zeta^2+\delta^2)^2\nonumber \\
\nonumber
&-&k_2(\frac{\sigma^4}{2}+\frac{\delta^4}{2}+3\sigma^2\delta^2+\zeta^4)\\
&-&k_3\chi(\sigma^2-\delta^2)\zeta\\&+&
k_4\chi^4+\frac{1}{4}\chi^4ln\frac{\chi^4}{\chi_0^4}-
\epsilon\chi^4ln(\frac{(\sigma^2-\delta^2)\zeta}{\sigma_0^2\zeta_0})
\nonumber
\end{eqnarray}

\noindent where $k_0$ ... $k_4$ are numerical constants, and the
vector meson self interaction is given by

\begin{eqnarray}
\mathcal{L}_{Vec}&=&-\frac{\chi^2}{2\chi_0^2}(m_\omega^2\omega^2+m_\rho^2\rho^2)\nonumber\\
&-&g_4^4 (\omega^4+6\beta \omega^2 \rho^2 + \rho^4)+... \; .
\end{eqnarray}

\noindent Here we ignore the $\phi$ meson as we only want to
study the effect on nucleons (plus a single hyperon). The explicit chiral symmetry
breaking term is given by

\begin{eqnarray}
&L_{SB}= m_\pi^2 f_\pi\sigma+\left(\sqrt{2}m_k^
2f_k-\frac{1}{\sqrt{2}}m_\pi^ 2 f_\pi\right)\zeta.
\end{eqnarray}

In the case of the baryonic octet we use an f-type coupling
between the baryons and vector mesons, which yields coupling
strengths as given by quark counting rules, i.e. $g_{i\omega} =
(n^i_q-n^i_{\bar{q}}) g_{8}^V$\,, $g_{i\phi}   =
-(n^i_s-n^i_{\bar{s}}) \sqrt{2} g_{8}^V$\,,
where 
$g_8^V$ denotes the vector coupling of the baryon octet and $n^i$
the number of constituent quarks of species $i$ in a given hadron.
Here, we allow for some slight tuning of the isovector channel
($g_{N\rho}$) to optimize the reproduction of isotopic chains.

Before performing a full-scale study of all known nuclei, several
test nuclei are selected, six crucial parameters $g_{N\omega}$,
$g_{N\rho}$, $g_4^4$, $\epsilon$,
 $\beta$ and $r$ are varied and the resulting nuclear
binding energies are studied. These parameters are the
nucleon-$\omega$ coupling $g_{N\omega}$, the nucleon-$\rho$
coupling $g_{N\rho}$, $g_4^4$, which governs the self-interaction
of the vector mesons and $\epsilon$, which governs a logarithmic
term in the scalar meson self-interaction and is usually very
small. The other two parameters are $\beta$, which controls mixing
terms between the $\omega$ and $\rho$ meson, and the ratio r,
which controls how much of the mass of the vector meson is generated by the scalar fields (r=1). 
In this way a parametrization is identified that best describes
nuclear binding energies.

 The test nuclei used
for this approach are a general set from light to heavy nuclei,
containing $^{16}O, ^{40}Ca, ^{48}Ca, ^{58}Ni, ^{90}Zr, ^{116}Sn,
^{124}Sn \text{and} ^{208}Pb$, a set for moderately heavy nuclei,
containing $^{100}Sn, ^{110}Sn, ^{116}Sn, ^{120}Sn, ^{124}Sn,
^{130}Sn, ^{132}Sn \text{and} ^{134}Sn$, and a set for heavy
nuclei, containing $^{182}Pb, ^{186}Pb, ^{192}Pb, ^{202}Pb,
^{208}Pb, ^{212}Pb \text{and} ^{214}Pb$.

After a good parametrization was found, the whole even-even nuclei
from the Wang-Audi-Wapstra data \cite{AudiWapstra}, 
\cite{AudiWapstra2003}, \cite{Wang1}, \cite{Wang2}, \cite{Wang3} were calculated, leading to an average
error in the description of binding energies of $0.44\%$.  The
resulting values for the parameters are listed in table
\ref{TS2ParameterTable}

\begin{table}[htb] \centering 
\begin{tabular}{|c|c|}
\hline Parameter          & Value  \\
\hline $g_{N\Omega}$   &   11.75       \\
\hline $g_{N\rho} $   &  4.385     \\
\hline $g_4^4$ &  38.75  \\
\hline $\epsilon$ & 0.0607   \\
\hline $\beta$ &  0   \\
\hline $r$ & 0.5 \\
\hline
\end{tabular}
\caption{Parameters of the TS2 model
parametrization.}\label{TS2ParameterTable}
\end{table}

\begin{figure}
\begin{center}
\includegraphics[width=0.5\textwidth]{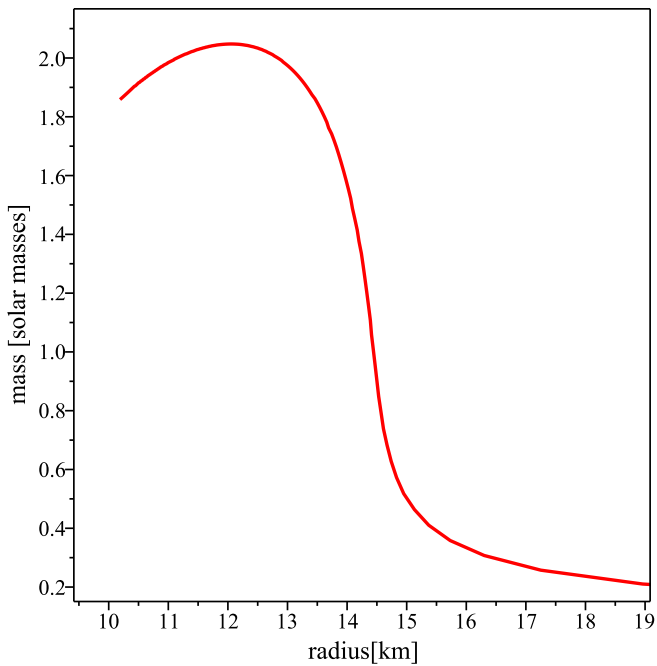}
\caption[Mass-Radius relation for neutron stars]{Resulting
mass-radius relation for neutron stars when applying the equation
of state.}\label{MRLarge}
\end{center}
\end{figure}

and shall hence be called TS2. Additionally, the model properties
were tested in other realms such as the equation of state with a
compressibility $\kappa =  275.13$\,MeV and hyperon
potentials at saturation density ($\Lambda$=-29.2 MeV, $\Sigma^+$=5.6 MeV, $\Xi^0$=-19.5 MeV).
 
A main motivation for refitting the model parameters was the
attempt to reproduce the masses of observed high-mass neutron
stars. These observations give valuable insight how the equation
of state behaves at very large densities. To check the star
masses, the Tolmann-Oppenheimer-Volkoff equations for spherical
stars was solved with this parametrization. The maximum mass of a
neutron star within this model is 2.05 $M_{\odot}$ at a radius of
12 km (see figure \ref{MRLarge}). The maximum central density reached in
the core is $\sim$ 5.5 $\rho_0$. As such, the model is in
accordance with current constraints on the neutron star
mass-radius relation.

\section{Results}

We will now discuss the results obtained for the dripline
calculations of the magnesium and magnesium + $\Lambda$
hypernucleus isotope chain. As a first test, known one-$\Lambda$
separation energies of already measured hypernuclei were checked.
We find that all three models give reasonable values about 0.5 to
1 MeV around the experimental values.

\begin{table}[htb] \centering 
\begin{tabular}{|c|c|c|c|c|}
\hline Nucleus           & Experiment & NL3    & SPL40       & TS2 \\
\hline $^4_{\Lambda}H$    &   2.04   &  2.116    & 2.729   & 2.179     \\
\hline $^9_{\Lambda}Li$    &  8.50   &  9.077   &  9.683   & 8.209    \\
\hline $^{17}_{\Lambda}O$ & 13.59 & 12.688 & 12.854 & 12.260 \\
\hline $^{41}_{\Lambda}Ca$ & 19.24 & 19.053 & 18.542 & 17.901 \\
\hline $^{56}_{\Lambda}Fe$ & 21.00 & 22.356 & 22.356 & 20.362 \\
\hline $^{139}_{\Lambda}La$ & 23.8 & 24.950 & 24.202 & 24.184 \\
\hline $^{208}_{\Lambda}Pb$ & 26.5 & 26.310 & 25.183 & 25.432 \\
\hline
\end{tabular}
\caption{$\Lambda$ separation energies in MeV for various nuclei
and models tested.
 For the experimental values, see \cite{SamantaMass1} 
 and references therein. The calculated
results are from the NL3, SPL40 and TS2
models.}\label{HyperLambdaTable}
\end{table}

Please note that there exist both conventions in the literature,
A=N+Z and A=N+Z+Y, with A denoting the nucleon or baryon number, respectively. We use the former one, A=N+Z, where the
$\Lambda$ is mentioned, but not counted.

For the dripline calculations, we use a two-dimensional code
\cite{SWSDeformedNuclei} to solve for the binding energy as a
function of the deformation of the nucleus. It is possible that
several minima arise at prolate or oblate deformation, and the
deepest minimum represents the energetically most stable
configuration of the nucleus within the model approximations.

As long as the drip line is not reached, the addition of a neutron
increases the binding energy of a nucleus but makes it more and
more unstable. When the dripline is reached, the system cannot
gain any energetic advantage anymore from adding another neutron.
Hence the binding energy of the first unbound nucleus will be
above the minimum of the binding energy of the last bound nucleus.
In this way, the dripline can be calculated, which is analogous to
looking for a sign change of the neutron separation energy
$S_{2n}$ (2n in this case as we only consider even-even nuclei).

\begin{figure}
\includegraphics[width=0.5\textwidth]{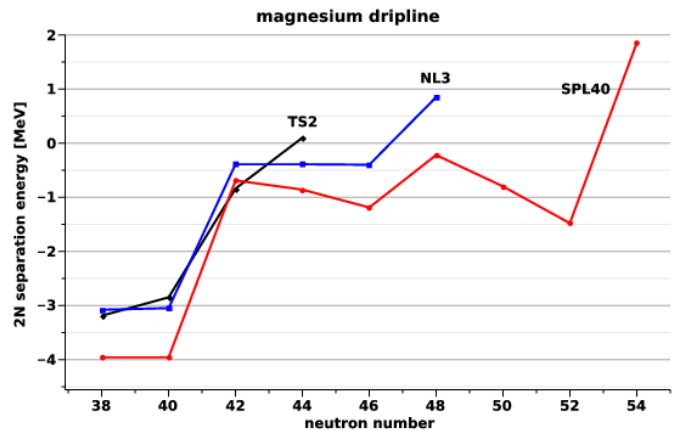}
\caption[$S_{2n}$ for magnesium]{2-neutron separation energy
$S_{2n}$for the magnesium chain within the various models. As long
as the energy is negative, it means that the next nucleus is more
deeply bound. A sign chain indicates instability with respect to
neutron emission. }\label{figure2}
\end{figure}

\begin{figure}
\includegraphics[width=0.5\textwidth]{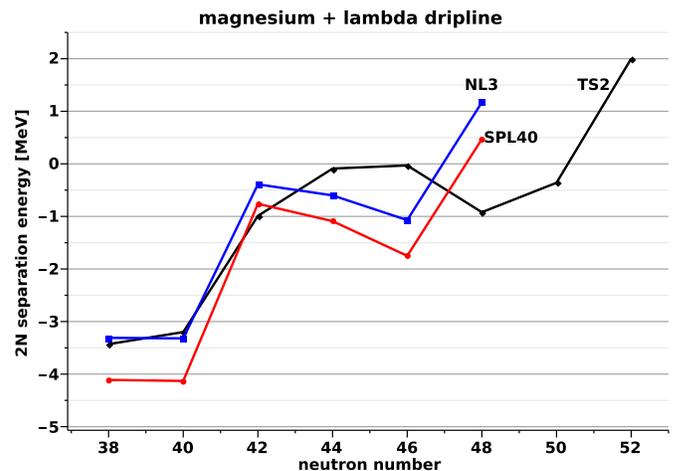}
\caption[$S_{2n}$ for magnesium+$\Lambda$]{2-neutron separation
energy $S_{2n}$ for the magnesium chain plus a $\Lambda$ hyperon
added for the various models. }\label{figure3}
\end{figure}

Figures \ref{figure2} and \ref{figure3} show the sign change and hence, the appearance of
the dripline, for the magnesium and magnesium + $\Lambda$ isotope
chain. Within our model, we can also calculate the binding energy
as a function of a specific deformation. This gives additional
insight into the microscopic structure of the nucleus, and differs
from simple liquid drop type calculations.

\begin{figure}
\begin{center}
\includegraphics[width=0.5\textwidth]{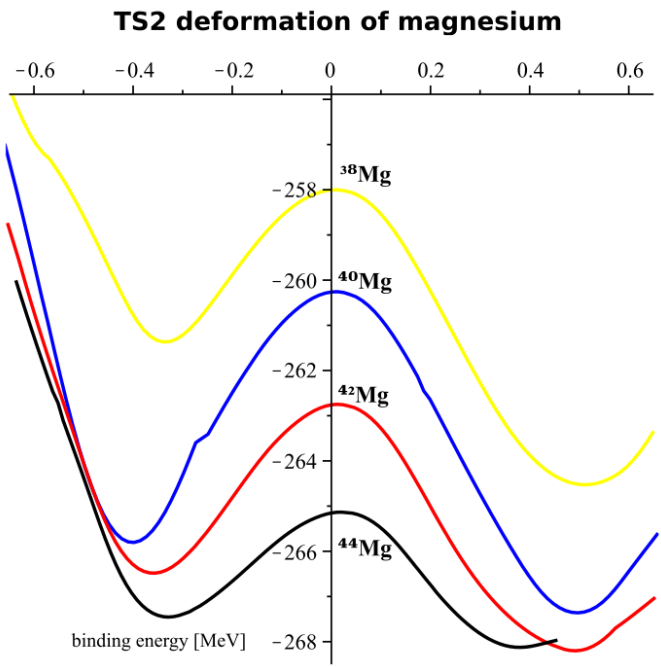}
\caption[Dripline for Mg in the NewBestFit parameter set]{Dripline
calculation for Mg in the TS2 parameter set. The dripline is at
$^{42}Mg$, as by a slight margin it is deeper bound than
$^{44}Mg$. Plotted is the deformation on the x axis and the total
binding energy in [MeV] on the y axis.}\label{figure4}
\end{center}
\end{figure}

\begin{figure}
\begin{center}
\includegraphics[width=0.5\textwidth]{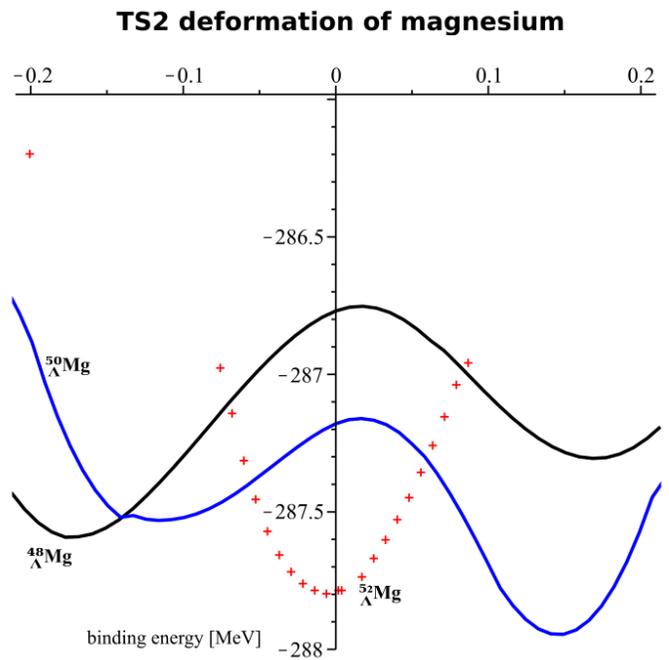}
\caption[Dripline for Mg with $\Lambda$ in the NewBestfit
parameter set.]{Dripline for Mg with $\Lambda$ in the TS2
parameter set. The dripline is at $^{50}_{\Lambda}Mg$. Plotted is
the deformation on the x axis and the total binding energy in
[MeV] on the y axis.}\label{figure5}
\end{center}
\end{figure}

Overall, the general structure of the binding energy as function
of deformation with resulting competing oblate and prolate minima
is in accordance with other calculations
\cite{Lalazissis:1997mb,RodriguezGuzman:2002nz,SWSDeformedNuclei}.

Figures \ref{figure4} and \ref{figure5} show the important portion of the calculation
where the dripline is reached for both the Mg and Mg + $\Lambda$
isotope chain. It is noteworthy that the last experimentally
measured nucleus of the Mg isotope chain is $^{40}Mg$
\cite{Mg40}. In the TS2 model, we find that the dripline is at
$^{42}Mg$. If a $\Lambda$ hyperon is added, this opens up an
additional degree of freedom, allowing for a wider energy
distribution and the addition of eight additional neutrons,
leading to a dripline at $^{50}_{\Lambda}Mg$. The general behavior
of additional binding is expected in hypernuclei, however the size
of the shift is quite remarkable.

When redoing the calculations for the SPL-40 and NL-3 parameter
sets, the driplines are found to be $^{52}Mg$ and $^{46}_{\Lambda}
Mg$ for SPL-40 and $^{46}Mg$ and $^{46}_{\Lambda}Mg$ for the NL3
parameter set.

\section{Discussion and conclusion}

The calculations of the stability region of Mg and Mg+$\Lambda$
show significantly different results for various models. All models
predict the $\Lambda$ separation energies of known hypernuclei
with good precision, see table \ref{HyperLambdaTable}.

When it comes to extrapolating to not yet experimentally measured
hypernuclei, results become very model dependent. The SPL-40
parameter shows a rather interesting behavior.
 In contrast to the
common conception that the addition of a hyperon does open up a
new energy pool and hence, allows more neutrons to be present up
to the fermi surface, here the opposite is the case. Instead, the
system becomes unstable and, if created, would eject six neutrons
to again reach the drip line.

Such a behavior would be very interesting. However, we do not
observe similar results with other models. Both
the NL3 and TS2 model confirm the general assumption that the addition
of a $\Lambda$ makes the nucleus more stable. However, all three
models give different results for the position of the dripline. In the NL3 model, 
the addition of a $\Lambda$ hyperon only changes the overall binding energy, but not the position
of the dripline. In the TS2 model, the dripline changes from $^{42}Mg$ to $^{50}_{\Lambda}Mg$
when a $\Lambda$ is added. When looking at the 2N separation energies, figure 3, the energy
becomes almost zero for the TS2 model around $^{44}_{\Lambda}Mg$ and $^{46}_{\Lambda}Mg$, but has 
a notable difference for $^{48}_{\Lambda}Mg$ again. This could hint that, depending on the resolution 
ability of the model, the dripline could also occur earlier, at $^{44}_{\Lambda}Mg$ or $^{46}_{\Lambda}Mg$.

Summarizing, in this paper we have presented a new parameterisation of a
chiral flavor-SU(3) Lagrangian that yields a good fit to nuclei as
well as reproducing 2-solar-mass neutron stars. Specifically, we
presented the isotope chain of Mg nuclei and the corresponding
hypernuclei and compared results with other relativistic
mean-field approaches. While all three models describe known
low-mass hyper-nuclear $\Lambda$ separation energies with good
precision, the extrapolation to more neutron-rich nuclei
remains theoretically challenging. Therefore, a main result of the study
is that such extrapolations to the dripline have found to be very model-dependent. As an
intriguing result we have further shown that the stability of
neutron-rich nuclei can be substantially modified by adding a
$\Lambda$-hyperon, possibly shifting the drip-lines considerably
by up to 8 neutrons.

Chhanda Samanta kindly acknowledges support provided by the
LOEWE-Program (HIC-FAIR).


\begin{thebibliography}{50}

\bibitem{Sugimura2014} H. Sugimura et al.,
\emph{Search for $^6_{\Lambda}H$ hypernucleus by the $^6$Li($\pi^-$, $K^+$) reaction at $p_{\pi^-}$ = 1.2 GeV/c},
 Phys. Lett. B \textbf{729}, 39 (2014); arXiv:1310.6104

\bibitem{Sugimura2013}Sugimura et al., \emph{Study of Neutron-Rich Hypernuclei by the ($\pi^-$, K+) Reaction at J-PARC},
Few Body Syst., \textbf{54}, 1235 (2013).

\bibitem{Gogami2013} T. Gogami et al, \emph{Electroproduction of K+$\Lambda$ at JLab Hall-C},
 Few Body Syst. \textbf{54}, 1227 (2013) 

\bibitem{Nakamura2011} S.N. Nakamura et al., \emph{Spectroscopic investigation 
of $\Lambda$ hypernuclei in the wide mass regction using the $(e,e’K^+)$ reaction},  
Jour. of Phys. \textbf{312}, 092047 (2011) 

\bibitem{HypHI} S. Bianchin et al., \emph{The HypHI project:
Hypernuclear spectroscopy with stable heavy ion beams and rare isotope beams at GSI and FAIR},
Int. J. Mod. Phys. E \textbf{18}, 2187 (2009); arXiv:0812.4148v1

\bibitem{HypH1Results} T.R. Saito et al.,  \emph{Latest Results From the HypHI Experiments at GSI: Hypernuclear
 Spectroscopy with Heavy Ion Induced Reactions}, Few-Body Syst. \textbf{54}, 1211 (2013).

\bibitem{Botvina} A. S. Botvina, \emph{Production of exotic hypernuclei and hyper-matter}, arXiv:1305.5474v1

\bibitem{NextDecade} Josef Pochodzalla, \emph{Hypernuclei - the next decade}, 
Acta Phys.Pol B \textbf{42}, 833 (2011);
arXiv:1101.2790v1

\bibitem{PL40Original} P.G.Reinhard, Z. Phys. A. \textbf{329}, 257 (1988).

\bibitem{PL40} K. Rutz, M. Bender, T. Bürvenich, T. Schilling,
P.-G. Reinhard, J. A. Maruhn, W. Greiner, \emph{Superheavy nuclei
in self-consistent nuclear calculations}, Phys. Rev. C \textbf{56}, 238 (1997).

\bibitem{NL3} G. A. Lalazissis, A.R. Farhan, M. M. Sharma, Nucl. Phys. A. \textbf{628}, 221 (1998).

\bibitem{Papazoglou 98a} Panajotis Papazoglou, Stefan Schramm, Jürgen
Schaffner-Bielich, Horst Stöcker, Walter Greiner, \emph{Chiral
Lagrangian for strange hadronic matter}, Phys. Rev. \textbf{57}, 2576 (1998).

\bibitem{Papazoglou Paper 98} P. Papazoglou, D. Zschiesche, S. Schramm, J. Schaffner-Bielich, H. Stöcker, W. Greiner,
\emph{Nuclei in a Chiral SU(3) model}, Phys. Rev. C \textbf{59}, 411 (1999); arXiv: nucl-th/9806087

\bibitem{Zschiesche2001} D. Zschiesche, P. Papazoglou, S. Schramm, J.Schaffner-Bielich, H. Stöcker, W. Greiner, \emph{Hadrons in
Dense Resonance-Matter: A Chiral SU(3) Approach}, Phys. Rev. C \textbf{63}, 025211 (2001); arXiv: nucl-th:0001055

\bibitem{Baryon Resonances Zschiesche} D. Zschiesche, G. Zeeb, S. Schramm, H. Stöcker, \emph{Impact of baryon resonances on the
chiral phase transition at finite temperatures and density}, J.Phys.G \textbf{31}, 935 (2005); arXiv:nucl-th/0407117v2

\bibitem{VeronicaHybrid} V. A. Dexheimer, S. Schramm, 
\emph{A novel approach to model hybrid stars}, Phys.Rev.C \textbf{81} 045201,(2010); arXiv:0901.1748v4

\bibitem{SWSDeformedNuclei} S. Schramm, \emph{Deformed nuclei in a chiral model},
Phys.Rev. C \textbf{66}, 064310 (2002); arXiv:nucl-th/0207060v1

\bibitem{Demorest} P. B. Demorest, T. Pennucci, S. M. Ramson, M. S. E. Roberts, J. W. T. Hessels, \emph{A two-solar mass neutron
star measured using Shapiro delay}, Nature,  \textbf{467}, 1081 (2010).

\bibitem{2Solar2013} John Antoniadis et al., \emph{A massive pulsar in a compact
relativistic binary}, Science \textbf{340} (2013)

\bibitem{AudiWapstra} G. Audi and A.H. Wapstra (1995),
 \emph{The 1995 Update to the Atomic Mass Evaluation}, Nuclear Physics A \textbf{595}, 409 (1995).

\bibitem{AudiWapstra2003} G. Audi, O. Bersillon, J. Blachot and A. H. Wapstra, Nucl. Phys. A \textbf{729}, 3 (2003);
 G.Audi, A.H.Wapstra and C.Thibault, Nucl. Phys. A \textbf{729}, 129 (2003);
 G.Audi, A.H.Wapstra and C.Thibault, Nucl. Phys. A \textbf{729}, 337 (2003).
  
\bibitem{Wang1} Audi, G., Kondev, F., Wang, M., Pfeiffer, B., Sun, X., Blachot, J., MacCormick, M.,
\emph{The NUBASE2012 evaluation of nuclear properties}, Chinese Physics C \textbf{36}, 1157 (2012).

\bibitem{Wang2} Audi, G., Wang, M., Wapstra, A., Kondev, F., MacCormick, M., Xu, X., Pfeiffer, B., 
\emph{The AME2012 atomic mass evaluation (I). Evaluation of input data, adjustment procedures},
Chinese Physics C \textbf{36}, 1287 (2012).

\bibitem{Wang3} Wang, M., Audi, G., Wapstra, A., Kondev, F., MacCormick, M., Xu, X., Pfeiffer, B.,
\emph{The AME2012 atomic mass evaluation (II). Tables, graphs and references}, Chinese Physics C,
\textbf{36}, 1603 (2012).

\bibitem{Lalazissis:1997mb}
  G.~A.~Lalazissis and P.~Ring,
  \emph{Excitation energy of superdeformed bands in relativistic mean field
  theory},
 Phys.Lett. B \textbf{427}, 225 (1998).


\bibitem{RodriguezGuzman:2002nz}
 R.~Rodriguez-Guzman, J.~L.~Egido and L.~M.~Robledo,
 \emph{Correlations beyond the mean field in magnesium isotopes: Angular momentum projection and configuration mixing},
 Nucl. Phys. A \textbf{709}, 201 (2002).

\bibitem{Mg40} T. Baumann, A. M. Amthor, D. Bazin, B. A. Brown, C.
M. Folden III, A. Gade, T. N. Ginter, M. Hausmann, M. Matos, D. J.
Morrissey, M. Portillo, A. Schiller, B. M. Sherrill, A. Stolz, O.
B. Tarasov, M. Thoennessen, \emph{Discovery of $^{40}Mg$ and
$^{42}Al$ suggests neutron drip-line slant towards heavier
isotopes}, Nature, \textbf{449}, 1022 (2007).

\bibitem{SWSRadii} S. Schramm, \emph{Nuclear and neutron star radii},
 Phys.Lett. B \textbf{560}, 164 (2003); arXiv:nucl-th/0210053

\bibitem{Veronica ProtoNS} Veronica Dexheimer, Stefan Schramm,
\emph{Proto-neutron and neutron stars in a chiral SU(3) model},
Astrophys.J. \textbf{683}, 943 (2008); arXiv:0802.1999

\bibitem{Schramm Zschiesche RotatingNS} S.Schramm, D. Zschiesche,
\emph{Rotating neutron stars in a chiral SU(3) model},
J.Phys.G \textbf{29}, 531 (2003); arXiv:nucl-th/0204075

\bibitem{Beckmann2} Ch. Beckmann, P. Papazoglou, D. Zschiesche,
S. Schramm, H. Stöcker, W. Greiner, \emph{Nuclei, super-heavy
nuclei and hypermatter in a chiral SU(3)-model}, Phys.Rev. C \textbf{65}, 024301 (2002) 024301; arXiv:nucl-th/0106014v1

\bibitem{TS Delta resonances} Torsten Schürhoff, Stefan Schramm,
\emph{Neutron stars with small radii -- the role of delta
resonances} Astrophysical Journal Letters \textbf{724}, L74 (2010).
[arXiv:1008.0957v1]

\bibitem{SamantaMass1} C. Samanta, P. Roy Chowdhury, D.N. Basu,
\emph{Generalized mass formula for non-strange and hyper nuclei
with SU(6) symmetry breaking}, J.Phys.G \textbf{32}, 363 (2006);
arXiv:nucl-th/0504085v2

\bibitem{SamantaMass2} C. Samanta, \emph{Generalized mass formula for non-strange,
strange and multiply-strange nuclear systems},
J. Phys. G: Nucl. Part. Phys. \textbf{37}, 075104 (2010); arXiv:1003.4227v1

\bibitem{Al23} A. Banu, L. Trache et al., \emph{Structure of $^{23}Al$ from
the one-proton breakup reaction and astrophysical implications},
Phys.Rev.C \textbf{84}, 015803 (2011); arXiv:1104.0675v2

\bibitem{SamantaCollaboration} H. Sugimura, M.Agnello, C. Samanta et al. (J-PARC E10 Collaboration), {Search for $^6_{\Lambda}H$ hypernucleus
by the $^6Li$ $(\Pi^-, K^+)$ reaction at $p_{\pi^-}=1.2 GeV/c$, Phys. Let. B \textbf{729}, 39 (2014).

\bibitem{O16} M. Ukai, S. Ajimura et al., {$\gamma$-ray spectroscopy of $^{16}_{\lambda}0$
and $^{15}_{\lambda}N$ hypernuclei via the
$^{16}0(K^-,\pi\gamma)$ reaction} , Phys. Rev. C \textbf{77}, 054315 (2008).

\bibitem{O16Lambda} R. H. Dalitz, D. H. Davis, T. Motoba, D. N.
Tovee, \emph{Proton emitting $\lambda$ states of
$^{16}_{\lambda}0^{*}$}, Nuclear Physics A \textbf{625}, 71 (1997). 

\bibitem{Quark Mean Field China} J. N. Hu, A. Li, H. Shen, H. Toki, \emph{Quark mean
field model for single and double $\Lambda$ and $\Xi$ hypernuclei}, arXiv:1310.3602v1

\bibitem{Jan Production} J. Steinheimer, K.  Gudima, A. Botvina, I. Mishustin, M. Bleicher, H. Stöcker,
\emph{Hypernuclei, dibaryon and antinuclei production in high energy heavy ion collisions: Thermal production
vs. Coalescence}, Physics Letters B \textbf{714}, 85, (2012); arXiv:1203.2547v2

\bibitem{LambdaLambda in NS} C. Albertus, J. E. Amaro, J. Nieves, \emph{$\Lambda\Lambda$ interaction and hypernuclei},
arXiv:1309.1484v1

\bibitem{PointCouplingModel} Y. Tanimura, K. Hagino, \emph{Description of single-$\Lambda$ hypernuclei
with relativistic point coupling model}, arXiv:1111.1488v1

}

\end{thebibliography}
\end{document}